\documentclass[twoside,onecolumnm,12pt]{article}

\usepackage{graphicx}
\usepackage{amsmath}
\usepackage{amssymb}

\usepackage[T1]{fontenc}                    
\usepackage[bitstream-charter]{mathdesign}
\usepackage[english]{babel}                 

\usepackage[a4paper,left=1.5cm,right=1.5cm,top=2.0cm,bottom=2.5cm]{geometry} 
\usepackage{lettrine}                       

\usepackage[hang, small, labelfont=bf, up, textfont=it, up]{caption}         
\usepackage{booktabs}                       

\usepackage{titling}                        
\usepackage[noblocks]{authblk}
\usepackage{hyperref}                       
\usepackage{xcolor}

\usepackage{abstract}                       

\newcommand{\upcite}[1]{\textsuperscript{\textsuperscript{\cite{#1}}}}

\newcommand{\Teff}{$T_{\rm eff}$}
\newcommand{\logg}{$\log g$}
\newcommand{\logggaia}{$\log g_{\rm Gaia}$}

\newcommand{\Vt}  {$\xi_t$}

\newcommand{\kms}{km\,s$^{-1}$}

\newcommand{\Eexc}{$E_{\rm exc}$}
\newcommand{\eps}[1]{$\log\varepsilon_{\rm #1}$}

\newcommand{\starname}{TYC\,$429$-$2097$-$1$}
\newcommand{\ciso}{$^{12}$C/$^{13}$C}
\newcommand\ion[2]{#1$\;${\scriptsize\rmfamily\uppercase\expandafter{\romannumeral#2}}\relax}

\pretitle{\begin{center}\Huge} 
\posttitle{\end{center}} 

\title{\textbf{The nature of the Li enrichment in the most Li-rich giant star}} 
\author[1,2]{\small Hong-Liang Yan}
\author[1,2]{{Jian-Rong Shi}\thanks{sjr@nao.cas.cn}}
\author[1,2]{Yu-Tao Zhou}
\author[3]{Yong-Shou Chen}
\author[4]{Er-Tao Li}
\author[5]{Suyalatu Zhang}
\author[6]{Shao-Lan Bi}
\author[6]{Ya-Qian Wu}
\author[3]{Zhi-Hong Li}
\author[3]{Bing Guo}
\author[3]{Wei-Ping Liu}
\author[1,2]{Qi Gao}
\author[1]{Jun-Bo Zhang}
\author[1,2]{Ze-Ming Zhou}
\author[1]{Hai-Ning Li}
\author[1,2]{{Gang Zhao}\thanks{gzhao@nao.cas.cn}}

\affil[1]{\small Key Laboratory of Optical Astronomy, National Astronomical Observatories, Chinese Academy of Sciences, Beijing 100012, China}
\affil[2]{School of Astronomy and Space Science, University of Chinese Academy of Sciences, Beijing 100049, China}
\affil[3]{China Institute of Atomic Energy, Beijing 102413, China}
\affil[4]{College of Physics and Energy, Shenzhen University, Shenzhen 518060, China}
\affil[5]{College of Physics and Electronics Information, Inner Mongolia University for Nationalities 028000, China}
\affil[6]{Department of Astronomy, Beijing Normal University, Beijing 100875, P. R. China}

\date{} 

\begin{document}
\maketitle


\noindent\textbf{About one percent of giants\upcite{Brown1989} are
detected to have anomalously high lithium (Li) abundances in their
atmospheres, conflicting directly with the prediction of the
standard stellar evolution models\upcite{Iben1967}, and making the
production and evolution of Li more intriguing, not only in the
sense of the Big Bang nucleosynthesis\upcite{Cyburt2016,Spite1982}
or the Galactic medium\upcite{Tajitsu2015}, but also the evolution
of stars. Decades of efforts have been put into
explaining why such outliers exist\upcite{Sackmann1999, Denissenkov2004,
Charbonnel2010}, yet the origins of Li-rich giants are still being
debated. Here we report the discovery of the most Li-rich giant
known to date, with a super-high Li abundance of 4.51. 
This rare phenomenon was snapshotted together with another
short-term event that the star is experiencing its
luminosity bump on the red giant branch. Such high Li
abundance indicates that the star might be at the very beginning of
its Li-rich phase, which provides a great opportunity to
investigate the origin and evolution of Li in the
Galaxy. A detailed nuclear simulation is presented with up-to-date
reaction rates to recreate the Li enriching process in
this star. Our results provide tight
constraints on both observational and theoretical points of view,
suggesting that low-mass giants can produce Li inside
themselves to a super high level via $^{7}$Be
transportation during the red giant phase.}


\vspace{10pt}
Lithium is too fragile to survive in deeper layers of a stellar
atmosphere due to the high temperature. Thus the
first dredge up (FDU) process can sharply dilute the surface Li
abundance in red giants. That explains why the
first discovery\upcite{Wallerstein1982} of a Li-rich giant evoked
great interests on exploring and understanding the Li-rich objects.
However, only about 150 Li-rich giants have been
found\upcite{Brown1989, Monaco2011, Kirby2012, Martell2013,
Adamow2014, Casey2016} in the past three decades, and $\sim 20$ of
them were found to be super Li-rich with
Li abundances higher than 3.3. Considering the NLTE
corrections, three\upcite{Martell2013, Balachandran2000, Reddy2005}
stars were found to be at a level of A(Li) $>4.0$. Such rare objects
could provide a great opportunity to reveal the
nature of the phenomenon of Li-richness because high Li abundance
cannot be maintained for a long time due to frequent
convection activity.
Taking advantage of the powerful ability for spectral collection
with the Large Sky Area Multi-Object Fiber Spectroscopy Telescope
(LAMOST), we have obtained a large sample of
Li-rich candidates by measuring the equivalent
width of the \ion{Li}{1} line at $\lambda=6707.8$\,\AA. One of our
candidates, \starname, has a super strong Li
absorption line (see Fig.~1, panel a). We
then made a follow-up high-resolution observation
with the $2.4$-m Automated Planet Finder Telescope (APF) located at
Lick Observatory on June 23, 2015. The spectrum covers a wavelength
range of $374$\,nm $-$ $970$\,nm with a resolution
of $\sim80,000$. The total integration time was 1.5 hours and was 
divided into three single exposures (30 minutes
each) for a better subtraction of cosmic-rays. The spectrum of
\starname\ obtained from APF is presented in Fig.~1, panels
(b) and (e), where the spectrum of HD\,$48381$ is also plotted with
a vertical shift of $+0.3$ as a comparison. HD\,$48381$ is a star
selected from the Gaia-ESO survey DR2, 
which has very similar stellar parameters to \starname.


We used the spectroscopic method to derive the stellar parameters
(see the `methods' section for details). 
We presented the final derived parameters
of \starname\ and the estimated errors in Table~1. 
The NLTE Li abundances for $6707.8$\,\AA, $6103.6$\,\AA, 
and $8126.3$\,\AA\ are $4.42\pm0.09$, $4.51\pm0.09$,
and $4.60\pm0.08$, respectively. The averaged Li abundance is
A(Li)$_{\rm{NLTE}}=4.51\pm0.09$. Compared to
previous studies, \starname\ has the highest Li abundance among
all Li-rich giants ever discovered (see Fig.~2). 
The Li abundance in \starname\ is about
$1,000$ times as high as the widely-used Li-rich `standard' of
A(Li)=1.5 (the lower purple dashed line in Fig.~2), despite
this `standard' being suggested to be
luminosity-dependent\upcite{Kirby2016}. It is also about 15 times as
high as meteoritic Li abundance (the upper purple
dashed line in Fig.~2), which is thought to be the initial
Li abundance for newly-formed young stars.

Although Li-rich giants were reported at various stages, such as 
RGB and core He-burning phases\upcite{Silva2014}, the Li-rich 
phase is likely to be a short-term event.
An extremely Li-rich giant (possibly newly enriched) with rigorous 
investigation on its evolutionary stage would be definitely important.
The location of the star was derived by the maximum
likelihood method using the observed parameters (in
this case, \Teff, \logg, and [Fe/H] derived from the spectroscopic
method) and a grid of evolutionary models computed
with the MESA code (see the `method' section for details). 
The derived luminosity and mass are
$\log (L/L_{\odot})$=$1.95$ and $M$=$1.43$\,$M_{\odot}$,
respectively. 
We used the parallax of Gaia DR1\upcite{Gaia2016} to test the reliability of the
information derived from the maximum likelihood method independently.
The luminosity obtained from Gaia data leads to a very 
similar result of $\log (L_{\rm{Gaia}}/L_{\odot})$=$2.00$.
The mass was tested in the sense that if the mass is well determined, the surface
gravity from Gaia parallax will show good consistency with the
spectroscopic \logg\ of $2.25$. As expected, the final result is
\logggaia$=2.23$. 
Thus, we consider that the results derived from the maximum
likelihood method are reliable, allowing us to robustly locate this
star on the Hertzsprung-Russel diagram (H-R diagram) along with the
corresponding MESA tracks (see Supplementary Figure 1). The star is 
likely occupying the region of the RGB-bump, a stage in which 
the $\mu$-barrier is destroyed and the enhanced
`extra-mixing' might be ongoing inside the star. In addition, we
also estimated the \ciso\ ratio as it has been suggested that the
extra mixing will cause a decrease of \ciso\ to the range of $10 -
20$. We found that the \ciso\ ratio in this star is
$12.0\pm3.0$, which is well within the predicted range. All the
results obtained above are shown in Table~1.

 \begin{table}[!t]
 \begin{center}
 \caption{The key information of \starname}\label{tab1}
 \begin{tabular}{rcl}
 \hline\hline\noalign{\smallskip}
   Property & & Value\\
 \noalign{\smallskip}
 \hline\noalign{\smallskip}
   Name                             &                & \starname       \\
   \Teff                            & (K)            & $4696\pm80$     \\
   \logg                            &                & $2.25\pm0.10$   \\
   $[$Fe/H$]$                       &                & $-0.36\pm0.06$  \\
   \Vt                              & (\kms)         & $2.30\pm0.10$   \\
   A(Li)$_{\rm NLTE}$               &                & $4.51\pm0.09$   \\
   \noalign{\smallskip}\hline\noalign{\smallskip}
   Gaia parallax $\pi$              & (milli-arcsec) & $0.73\pm0.24$   \\
   Mass                             & ($M_\odot$)    & $1.43\ ^{+0.55}_{-0.54}$ \\
   \noalign{\smallskip}
   $\log (L/L_{\odot})$             &                & $1.95\ ^{+0.25}_{-0.19}$ \\
   $\log (L_{\rm{Gaia}}/L_{\odot})$ &                & $2.00\pm0.06$   \\
   \logggaia                        &                & $2.23\pm0.16$   \\
   \noalign{\smallskip}\hline\noalign{\smallskip}
   \ciso                            &                & $12.0\pm3.0$    \\
   $v\sin i$                        & (\kms)         & $11.3\pm1.5$    \\
   $[\alpha$/Fe$]$                  &                & $0.19\pm0.04$   \\
 \noalign{\smallskip} \hline
 \end{tabular}
 \end{center}
 \end{table}


It has long been suggested that the Li enrichment could be due to
contaminations by external sources in the
environment, such as the engulfment of a substellar
component\upcite{Alexander1967} (e.g. giant planets or brown dwarfs)
and the accretion from a Li-rich companion or
diffuse medium. Yet the contribution from external sources
is not infinite, since the contributor itself has a
limited amount of Li, typically not higher than 3.3. A simulation on
engulfment of a Jovian planet suggested
that a typical upper limit for enrichment by such
way is $\sim 2.2$\upcite{Aguilera2016}. Our star has
a much higher Li abundance than any of those
values, thus it is very unlikely that the
overabundant Li comes from the direct contribution of external
sources.

The internal production of Li, on the other hand, is based on the
Cameron$-$Fowler mechanism\upcite{Cameron1971} (CF mechanism). The
production of $^{7}$Be takes place where the temperature is too high
to preserve the newly synthesized $^{7}$Li, hence $^{7}$Be must be
transported quickly to the cooler region to form
Li. This scenario would potentially require the
low-mass giants to evolve to the RGB-bump, where
the mean molecular weight discontinuity (or $\mu$-barrier, a mass
gradient caused by FDU) is erased. Meanwhile, it would need the
presence of deep, enhanced `extra mixing' (EM) to increase the depth
and efficiency of the convective circulation, which in-turn alters
the \ciso\ into a lower level than that after FDU. The observational
features on both the evolutionary stage and \ciso\ ratio of our star
coincide with these predictions remarkably well,
but the limitation of self-production still remains unknown in the
sense that none of the quantitative calculations
with a nuclear reaction network has been presented to obtain such
high amount of Li before. To test this speculation, we have made
such simulation with a series of parameters. By using the RGB
stellar structure as the input for the EM calculation, with the
updated nuclear reaction rates and the asymmetric parameters of the
EM model, we found that A(Li) in the envelope can
exceed $4.0$ for the processed material when the mass circulation
finished. Our EM calculation with parameters of $\dot{M} = 52$,
$\Delta$=0.15, $f_d$=0.9 and $f_u$=0.1 (see the `methods' section
for the details) yields A(Li)$=4.506$, where $\dot{M}$ is the rate of mass transport in units of 10$^{-6}$ M$_{\odot}$ yr$^{-1}$, $\Delta$ is $\log T_H-\log T_p$, where $T_H$ is the temperature at which the energy released from the H-burning shell reaches maximum and $T_p$ is the maximum temperature sampled by the circulating material, and $f_d$ and $f_u$ are the fractional areas of the `pipes' occupied by the mass flows moving downward and upward, respectively, and their values satisfy $f_d+f_u$=1. This reproduces the observed Li abundance for \starname well. Repeating the same
calculation with the alternative set of nuclear reaction rates from
the JINA database\upcite{Cyburt2010} yields a similar abundance of
A(Li)$= 4.515$. As a contrast, assuming this star had never experienced any EM, 
the Li abundance would be constant at the initial value of A(Li)=1.16, 
because the temperature in the envelope is too low to ignite both the 
production and destruction reactions of $^7$Li. The abundances of 
$^3$He, $^7$Li and $^7$Be as functions of the processing time for the 
mass circulation are shown in Fig.~3.

During the EM process, $^3$He is converted to $^7$Be via the
reaction of $^3$He($^4$He, $\gamma$)$^7$Be, and then $^7$Be is
quickly converted to $^7$Li via the reaction of $^7$Be(e$^-$,
$\nu$)$^7$Li.
To achieve such high level of Li abundance, abundant $^3$He is
required. The initial surface $^3$He is computed from the MESA
model, which is Y($^3$He) = $4.038\times10^{-4}$. Fig.~3
shows the decrease of $^3$He as a function of the time for EM
processing. A total amount of Y($^3$He)/H $\sim1.477\times10^{-4}$ is
burned-off during this circulation, and the produced Y($^7$Li)/H is
$3.206\times10^{-8}$. This is because another reaction,
$^3$He($^3$He, $2p$)$^4$He, dominates over the reaction
$^3$He($^4$He, $\gamma$)$^7$Be, thus consuming the majority of
$^3$He. The strong competition from $^3$He($^3$He, $2p$)$^4$He
reaction prevents more $^3$He from converting to $^7$Li. Testing
with different sets of EM parameters shows that the maximum of Li
abundance from our network calculation is $5.07$.
The $^3$He supply may eventually run out and cannot be renewed by
the giant, then the surface Li abundance is likely to decrease, even
if the internal conditions remain the same. On the other hand, if
the internal conditions do change, the surface Li abundance may also
decline due to the destruction by convective activities in stars.
Either way, the super Li-rich phase may disappear after a short
period of time.
In our calculation, the \emph{asymmetric} mass circulation described
by a large ratio of $f_d/f_u$ is a key factor for achieving super
high Li-enrichment. This large $f_d/f_u$ ratio indicates that the
upward flow is moving much faster (since its `pipe' is thinner) than
the downward flow, while the mass is conservative in the EM process.

The cause of the EM has not been well understood, and rotationally
induced mixing is often attempted. Indeed, \starname\ is a slightly
rapid rotator with a projective velocity of
$11.3$\,\kms, which is about ten times faster than that for normal
giants. The spinning up of an RGB star is either caused by the tidal
synchronization effects in a close binary system or the engulfment
of a massive planet\upcite{Denissenkov2004, Alexander1967}. We
calculated the radial velocities based on the two independent
observations through LAMOST and APF (with an interval of 10 months),
and found no significant radial velocity change at a level of a few
kilometers per second, which is the typical uncertainty for the RV
derived from LAMOST spectra. Thus it is very unlikely that a star
has a stellar companion which is massive and close enough to spin up
via tidal synchronization. On the other hand, one would expect some
associated features that are detectable if a massive planet was
engulfed and digested. For example, it was found that there might be
a large probability for Li-rich giants exhibiting excess in the
infrared (IR) flux, yet we found no sign of IR excess 
(see Supplementary Figure 2).
In addition, if the matter exchange did happen at a certain time, 
there should be some fluctuations in the abundance
pattern. However, \starname's \ciso\ is at a
typical level for its stage\upcite{Denissenkov2004}, and its
$\alpha$-abundance is also quite normal among the giants with
similar [Fe/H]. Given all these facts, we speculate that in our
case, the enhanced extra mixing might neither be caused by the
presence of a massive planet (if there were any) nor a close stellar
companion. There are other assumptions often approximated as the
\emph{internal} cause of enhanced extra mixing, i.e., thermohaline
instabilities and magnetic buoyancy. The thermohaline convection
driven by the $^3$He$(^3$He,$2p)^4$He reaction which produces a
local depression in the mean molecular weight\upcite{Charbonnel2010}
can cause enhanced extra mixing inside the star.
The magnetic buoyancy mechanism in the presence of a magnetic dynamo
would permit the buoyancy of magnetized material near the H-burning
shell, thus inducing the form of matter circulation in RGB
stars\upcite{Busso2007}. We speculate that the magnetic buoyancy and thermohaline instabilities might play roles together during the mass circulation, in which the former may lead to very fast upward circulation and the latter drive downward circulation at a much slower speed.

Although the Li abundance measured in this star is super high, it is
still well within the upper-limit that the theoretical model could
reach. It is also important to note that the RGB-bump is not the
only stage for inhabitation of Li-rich giants, many Li-rich giants 
have been reported in various stages by previous
work, including the core He-burning phase, which is very close to
the RGB-bump region on the H-R diagram. Although it is not preferred
by our data, if our star occupies this stage, a new scenario will be
in urgent need for interpreting such high Li abundance.


\small

\noindent \textbf{Acknowledgements} {This research was supported by National Key Basic Research Program of China 2014CB845700, the National key Research and Development Project of China 2016YFA0400502, and the National Natural Science Foundation of China under grant Nos. 11390371, 11603037, 11473033, 11490560, 11505117, 11573032, 11605097. Guoshoujing Telescope (LAMOST) is a National Major Scientific Project built by the Chinese Academy of Sciences. Funding for the project has been provided by the National Development and Reform Commission. LAMOST is operated and managed by the National Astronomical Observatories, Chinese Academy of Sciences. This work is supported by the Astronomical Big Data Joint Research Center, co-founded by the National Astronomical Observatories, Chinese Academy of Sciences and the Alibaba Cloud. This research uses data obtained through the Telescope Access Program (TAP). The authors acknowledge Dr. James Wicker for proof-reading the manuscript. We acknowledge the use of {\it Gaia} and WISE data, and of VizieR catalogue access tool. 
}

\vspace{10pt}
\noindent \textbf{Author contributions} {H.-L.Y., J.-R.S. and G.Z. proposed and designed this study. H.-L.Y and J.-R.S. led the data analysis with the contributions from Y.-T.Z., Q.G., J.-B.Z., and Z.-M.Z.. Y.-S.C., E.-T.L., S.Z., Z.-H.L., B.G., and W.-P.L. performed the nuclear calculations. S.-L.B. and Y.-Q.W. calculated the evolutionary models and tracks. H.-N.L. carried out the observation. All the authors discussed the results and contributed to the writing of the manuscript.
}

\vspace{10pt}
\noindent \textbf{Author Information} {Reprints and permissions information is available at www.nature.com/reprints. The authors declare no competing financial interests. Readers are welcome to comment on the online version of the paper. The data that support the plots within this paper and other findings of this study are available from the corresponding author upon reasonable request. Correspondence and requests for materials should be addressed to J.-R.Shi (sjr@nao.cas.cn) or G. Zhao (gzhao@nao.cas.cn). }

\clearpage
\small
\begin{figure*}[!h]
\setlength{\abovecaptionskip}{-15pt}
\setlength{\belowcaptionskip}{-15pt}
\begin{center}
\includegraphics[angle=0, width=15cm]{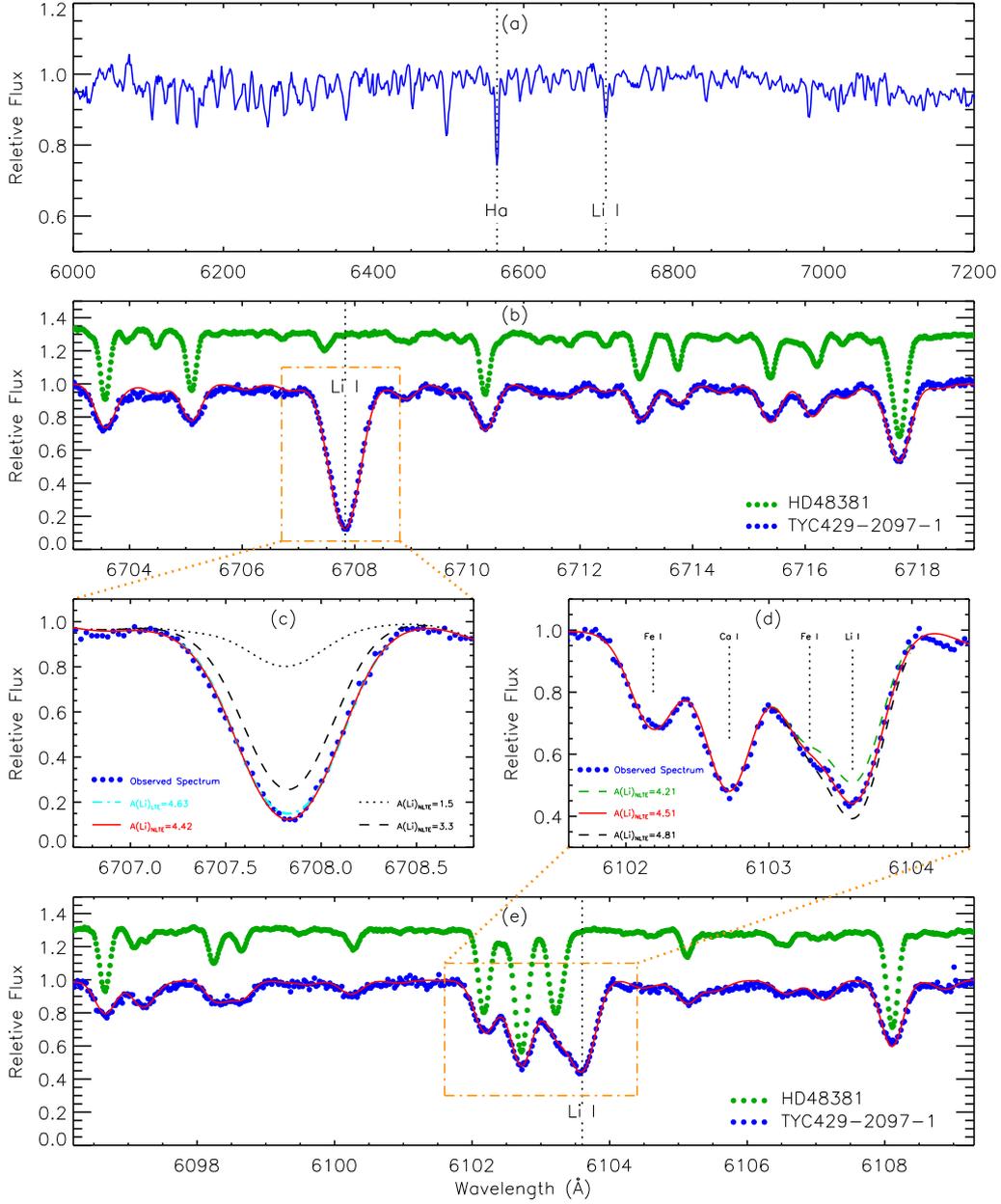}\\
\caption{The observed spectra and
line profile fittings for \starname. The
low-resolution spectrum from LAMOST is shown in panel (a), where
H$\alpha$ line and \ion{Li}{1} resonance
line are indicated. Panels (b) and (e)
display the high-resolution spectrum (blue dot)
observed by APF near 6708\,\AA\ and 6103\,\AA, respectively.
The spectrum of HD\,48381 is also plotted (green
dots) for comparison in both panel. The best
profile fittings for these two lines are indicated
with the red solid line in panels (c) and (d), respectively, where
several profiles computed with different Li abundances
are also presented. The profile from LTE
calculation reaches saturation (cyan curve) in panel (c), and
increasing Li abundance will barely affect the computed profile,
therefore the LTE abundance shown in panel (c) is estimated from the
equivalent width of this line.}\label{fig1}
\end{center}
\end{figure*}

\clearpage
\small
\begin{figure}[!h]
\setlength{\abovecaptionskip}{-25pt}
\setlength{\belowcaptionskip}{-15pt}
\includegraphics[angle=0, width=\hsize]{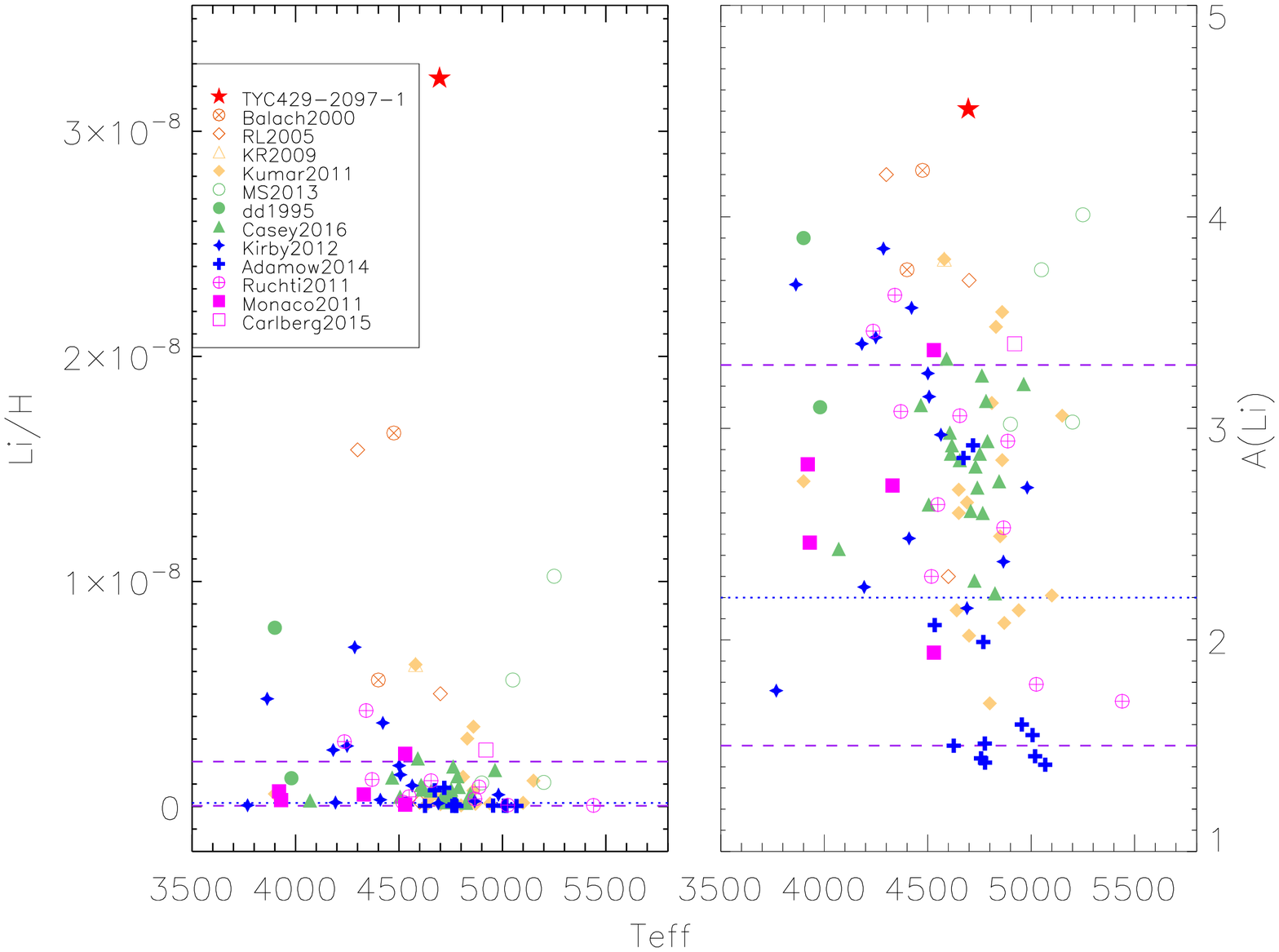}\\
\caption{The distribution of Li-rich giants in \Teff\
- Li/H plane (left panel) and \Teff\ - A(Li) plane (right panel).
\starname\ is indicated with a red star. The cited data are from
different literatures as illustrated in the left panel, namely:
Monaco2011\upcite{Monaco2011}, Kirby2012\upcite{Kirby2012},
MS2013\upcite{Martell2013}, Adam{\'o}w2014\upcite{Adamow2014},
Casey2016\upcite{Casey2016}, Balach2000\upcite{Balachandran2000}, 
RL2005\upcite{Reddy2005}, Kumar2011\upcite{Kumar2011},
dd1995\upcite{delaReza1995}, KR2009\upcite{Kumar2009},
Ruchti2011\upcite{Ruchti2011}, and
Carlberg2015\upcite{Carlberg2015}. All the Li abundances adopted
here are based on NLTE calculations. If the original work did not
perform the NLTE abundance analysis, we applied the NLTE corrections
interpolated from Lind's grid\upcite{Lind2009} to the original LTE
abundances. The horizontal dashed lines in purple indicate A(Li)=1.5
and A(Li)=3.3, respectively. The dotted blue line
shows the upper limit of Li enriched by engulfment
of a giant planet\upcite{Aguilera2016}.}\label{fig2}
\end{figure}

\clearpage
\small
\begin{figure}[!t]
\setlength{\abovecaptionskip}{-15pt}
\setlength{\belowcaptionskip}{-15pt}
\includegraphics[angle=0, width=\hsize]{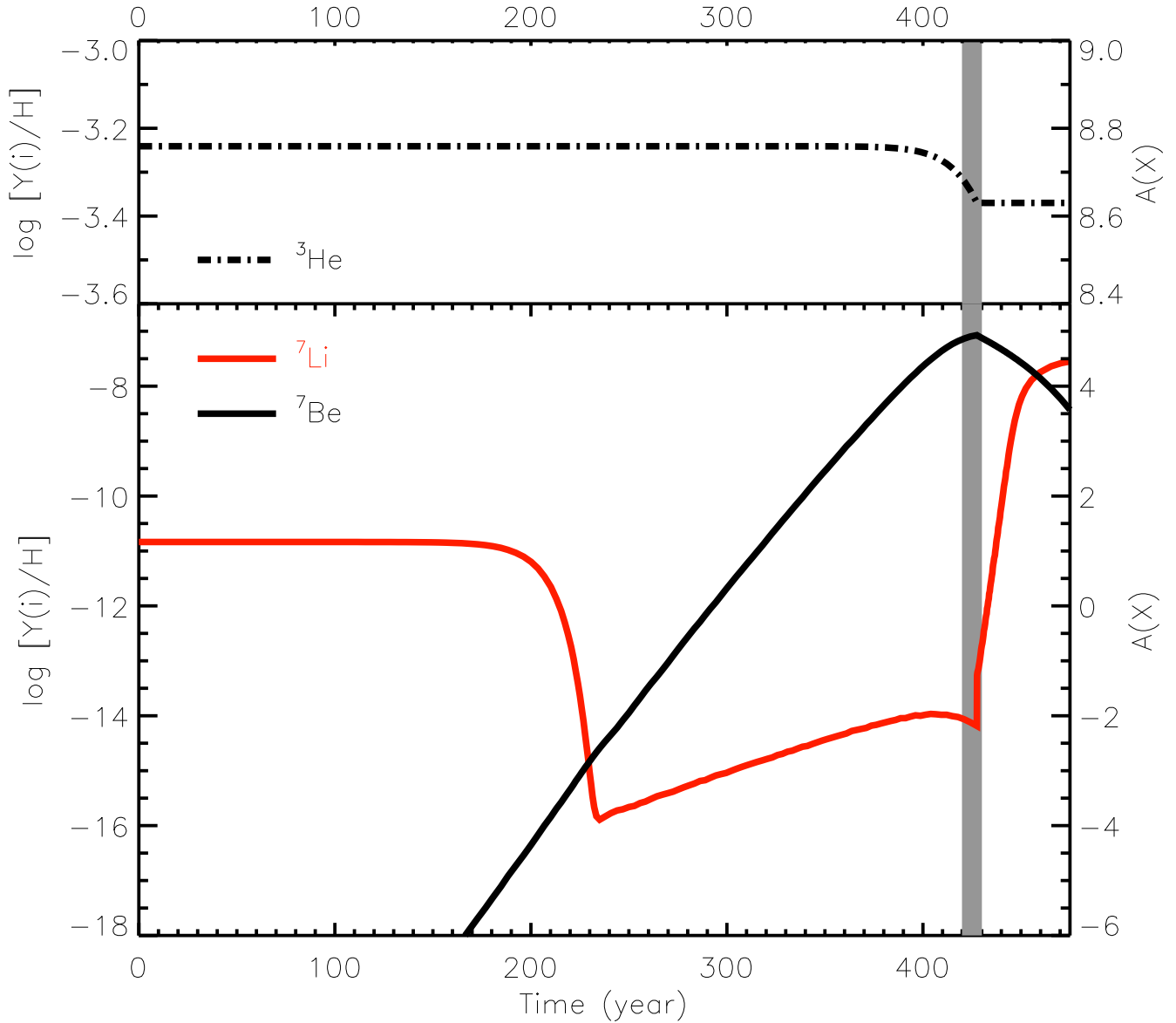}\\
\caption{The calculated surface abundances and mass
fraction of $^3$He, $^7$Be and $^7$Li as functions of the processing
time for the mass circulation. The left-hand y-axis
indicates the mass fractions in logarithmic scale and the right-hand
y-axis indicates the logarithmic abundances. The initial time is
set to be zero when the sample material at the base of the envelope
starts the mass circulation. The vertical solid line indicates the
boundary between the downward and upward motion of the processed
material. The yield of $^7$Be reaches maximum by
the end of the downward motion at t=$423$ yrs, and then $^7$Li 
begins to dominate during the upward motion of the
sample material for $47$ yrs. }\label{fig3}
\end{figure}

\clearpage
\appendix
\section*{Methods}

\noindent\textbf{Data Reduction.} We followed the standard procedure for data reduction with an Interactive Data Language (IDL) package, which was originally designed for the FOCES spectrograph\upcite{Pfeiffer1998}. The instrumental response and background scatter light were considered during the reduction, and cosmic rays and bad pixels were removed carefully. The resulting spectrum has a signal-to-noise ratio of $\sim 160$ at $6707.8$\,\AA.

\vspace{10pt}
\noindent\textbf{Deriving the Stellar Parameters.} We first combined three iron (Fe) line lists\upcite{Takeda2002, Mashonkina2011, Carlberg2012} and calibrated $213$ lines out of $257$ with the solar spectrum\upcite{Kurucz1984}. Then we eliminated those seriously blended or poorly recognized lines seen from the spectrum of \starname, as well as the lines that are too strong ($> 120$\,m\AA) or too weak ($< 20$\,m\AA). Finally, $57$ \ion{Fe}{1} and $12$ \ion{Fe}{2} lines are used as the parameter indicators. The effective temperature (\Teff) is derived from the excitation equilibrium of \ion{Fe}{1} lines with excitation energy (\Eexc) greater than $2.0$\,eV\upcite{Sitnova2015}. The surface gravity (\logg) is approached by equalizing the two sets of Fe abundances obtained from \ion{Fe}{1} and \ion{Fe}{2} lines, respectively. Statistically, the Fe abundance derived from each individual \ion{Fe}{1} line and the equivalent width (EW) from the same \ion{Fe}{1} line will be mutually independent if the micro-turbulence velocity (\Vt) is correctly set. Using this trick, we can obtain \Vt, and then the metallicity ([Fe/H]) can be settled simultaneously if all the mentioned constraints are achieved. All the Fe abundances are derived from NLTE analysis with the MARCS atmospheric models\upcite{Gustafsson2008} since it has been suggested that \ion{Fe}{1} lines suffer a non-negligible NLTE effect\upcite{Mashonkina2011}. The procedure of this approach is much more like an iteration. We started with the results from LAMOST pipeline as the initial input, and then by calculating MARCS models and adjusting the stellar parameters step by step, we finally end up with a self-consistent solution. Supplementary Figure 3 shows the derived Fe abundances from individual lines as functions of their EWs (upper panel) and \Eexc\ (lower panel). Based on the experience of our previous work using the similar spectroscopic method, the errors for \Teff, \logg, [Fe/H] and \Vt\ are estimated to be $\pm 80$\,K, $\pm 0.10$\,dex, $\pm0.06$\,dex and $\pm 0.10$\,\kms, respectively.

\vspace{10pt}
\noindent\textbf{Determination of the Elemental Abundances.} For all the species discussed in this paper, we use the spectrum synthesize method to derive their abundances. The theoretical profiles of the corresponding lines are calculated based on the MARCS model\upcite{Gustafsson2008}. An interactive IDL code Spectrum Investigation Utility (SIU) was applied to calculate the synthetic line profiles. The coupled radiative transfer and statistical equilibrium equations for the NLTE calculation were solved following the efficient method with a revised DETAIL program based on the accelerated lambda iteration, we refer readers to Mashonkina \emph{et al.} (2011) for a more detailed description of this method\upcite{Mashonkina2011}. The resulting departure files are transferred into SIU for NLTE line synthesize. The solar iron abundance of \eps{Fe} = 7.5 was assumed in our work.

In the abundance analysis of Li, the resonance line at $6707.8$\,\AA, the subordinate line at $6103.6$\,\AA\, and the line at $8126.3$\,\AA\upcite{Adamow2015} were used to derive the Li abundance. Although the line at $8126.3$\,\AA\ is blended with two telluric lines, it shows similar result to those derived from the resonance and subordinate lines. The final Li abundance is determined by averaging the results from these three lines. 
It is noted by many previous studies that the NLTE corrections are important for strong lines. In general, the NLTE correction for Li is not large for the `Li-normal' stars, however it will significantly increase for Li-rich objects, especially for the strong resonance line at $6707.8$\,\AA. In the very extreme cases (such as ours), the local thermodynamic equilibrium (LTE) theoretical profile of $6707.8$\,\AA\ could be saturated at the core. Therefore, the NLTE effects were taken into consideration in our abundance analysis for Li. 
For the NLTE analysis, we applied the same atomic model and line data as those presented in Shi \emph{et al.} 2007\upcite{Shi2007}. The carbon abundances were derived from the \ion{C}{1} line at $5086$\,\AA, and the C$_{\rm 2}$ line at $5135$\,\AA\upcite{Alexeeva2015}. The nitrogen lines were either blended or too weak in our spectrum, so we turned to the CN band near $8003.5$\,\AA\ to estimate the N abundance by fixing C to the value we just derived. Then the carbon isotopic ratio was determined by adjusting the contributions from $^{12}$C and $^{13}$C until we get the best fit to the CN band observed profile. The determination of $\alpha$-abundance (Mg, Si, and Ca) with NLTE analysis was based on a series of previous work\upcite{Mashonkina2013, Zhang2016, Mashonkina2007}. The final $\alpha$-abundance was obtained by averaging the abundances obtained from those elements. We also derived abundances of several other elements, and they can be found in Supplementary Table 1.

The error of the Li abundance was estimated by changing the stellar parameters (namely \Teff, \logg, and [Fe/H]) within their error ranges and calculating the corresponding variations on the abundance. The result of this test is presented in Supplementary Table 2. It is clear that the Li abundance is more sensitive to the variation of \Teff\ than that of \logg\ or [Fe/H]. A change of $80$\,K on \Teff\ will result in a variation of $\sim 0.09$\,dex for the Li abundance. Thus, we adopted the variation caused by the error of the effective temperature as the uncertainty for each Li I line, which is $\pm0.09$, $\pm0.09$, and $\pm0.08$ for the lines of $6707.8$\,\AA, $6103.6$\,\AA, and $8126.3$\,\AA\, respectively. The error of the final Li abundance is obtained by calculating the standard deviation of abundances derived from the three lines. For the other elements, if more than three lines are used for the abundance determination, we calculate their standard deviation and compare it to the error caused by the uncertainties of the stellar parameters, and the larger one was adopted as the final error. A few of species such as \ciso\ ratio were not suitable for the above analysis, we estimated their errors by giving the upper- and lower- limit of the best-fit to the profile.

\vspace{10pt}
\noindent\textbf{Maximum Likelihood Method and Evolutionary Stage.} The likelihood function is expressed following Basu \emph{et al.}\upcite{Basu2010}, which is defined as:
  \begin{equation}
  \label{e1}
    L=L_{P_{\rm{obs1}}}\ L_{P_{\rm{obs2}}}\ L_{P_{\rm{obs3}}}\ \cdot\ \cdot\ \cdot\ ,
  \end{equation}
where $P_{\rm{obs}}$ is the observed parameter (e.g., \Teff) and
  \begin{equation}
  \label{e2}
    L_{P_{\rm{obs}}} =\frac{1}{\sqrt{2\pi} \sigma_{P_{\rm{obs}}}}\exp\left(\frac{-(P_{\rm{obs}}-P_{\rm{model}})^2}{2\sigma_{P_{\rm{obs}}}^2}\right).
  \end{equation}
The normalized probability of each model $p_{i}$ is expressed as
  \begin{equation}
    p_{i}=\frac{L_{i}}{\sum_{i=1}^{N_{\rm m}}L_{i}},
  \end{equation}
where $L_{i}$ is the likelihood function of the $i$th model and $N_{\rm m}$ is the total amount of the models. The probability is integrated from the boundary constrained by 3$\sigma$ error range of the observed parameters. 
Thus, the maximum value of the integrated probability is 0.5, and the best-fitted parameters are obtained from this probability\upcite{Wu2017}.

The grid of evolutionary models for calculating the likelihood are generated from MESA\upcite{Paxton2011} code. The grid covers a wide-range of mass from $0.6$--$3.0$\,M$_{\odot}$ with a $0.02$\,M$_{\odot}$ interval on mass and a 0.005 interval on metallicity (Z). The evolution tracks are constructed from the pre-main sequence to the asymptotic giant branch (AGB) phase. For generating the grid, the initial parameter setup was mostly as same as described in Paxton \emph{et al.} 2011\upcite{Paxton2011} except for the solar chemical abundance $(Z/X)_{\odot}$. We adopted $(Z/X)_{\odot}=0.0229$\upcite{Grevesse1998} because calibrated solar model with this mixture fits the internal structures from helioseismic inversion\upcite{Bi2011} slightly better than the others\upcite{Asplund2009}. The MESA $\rho$-T tables are based on an updated version of Rogers \& Nayfonov tables\upcite{Rogers2002}. Ferguson \emph{et al.}\upcite{Ferguson2005} extended the opacity for the solar composition to the low-temperature case in 2015, and we adopted their results in our computation. The stellar metallicity was transferred into `metal' abundance $Z_{\rm{init}}$, from where the hydrogen and helium abundances ($X_{\rm{init}}$ and $Y_{\rm{init}}$) were calculated.

To obtain the luminosity and \logggaia\ from the parallax, we first calculated the bolometric magnitude for the absolute V magnitude by $M_{\rm bol}=V_{\rm mag}+BC+5\log\pi+5-A_{\rm V}$ (in this equation, $\pi$ is used in the unit of arcsec), where the bolometric correction BC was computed following Alonso \emph{et al.}\upcite{Alonso1999} and $A_{\rm V}$ was estimated by the Galactic extinction map presented by Schlafly \& Finkbeiner at 2011\upcite{Schlafly2011} (all values that were need for the calculation were presented in Supplementary Table). We calculated the luminosity using the relation of $\displaystyle M_{\rm bol}-M_{\rm bol\odot}=-2.5\log(\frac{L_{\rm{Gaia}}}{L_{\odot}})$. Finally, the \logggaia was determined following the fundamental relation $\displaystyle \log g = \log g_{\odot}+\log (\frac{M}{M_{\odot}})+4\log(\frac{T_{\rm eff}}{T_{\rm eff\odot}})+0.4(M_{\rm bol}-M_{\rm bol\odot})$, where $\log g_{\odot}=4.44$, $M=1.36\,M_{\odot}$, $T_{\rm eff\odot}=5777$\,K, and $M_{\rm bol\odot} = 4.74$\,mag. The errors were given via the uncertainty transfer formula assuming that the errors were contributed by the uncertainty of the Gaia parallax.

\vspace{10pt}
\noindent\textbf{Parametric Calculation of Internal Li-enriching.} During the RGB stage of a low-mass star, we introduced the `extra-mixing' or the `circulation process' after the H-burning shell has erased the chemical discontinuity left behind by FDU, signatured by the bump of the luminosity function in RGB. We followed the parameterization of Nollett \emph{et al.}\upcite{Nollett2003} to perform the parametric calculations using $\dot{M}$, $T_p$, $f_d$ and $f_u$ as free parameters.

The basic assumptions are described as follows. The EM is a process of meridional circulation in the radiative zone of low-mass red giant stars, the path of mass flow looks like a `conveyor belt'. A parcel of material with the initial abundance composition $Y^E_i(0)$ at the base of the convective envelope circulates downward through the radiative zone structure and finally returns to the envelope with newly processed abundance composition $Y^P_i$. The velocity of the sampled material can be expressed as $dr/dt=\dot{M}/f4\pi r^2\rho$, where $f$ is equal to $f_d$ (or $f_u$) for the downward (or upward) circulation, and the density $\rho$ is a function of radius $r$, governed by the stellar structure. The stellar structure, $\dot{M}$ and $\Delta$ specify as functions of time, the conditions of temperature $T$ and density $\rho$ where the material pass through.

The computing code integrates the network of reactions of the hydrogen burning chain by following the circulation trajectory. The nuclear reaction rates adopted are from NACRE \upcite{Angulo1999} except for two rates, the updated rate of $^7$Be$(p,\gamma)^8$B is from Du \emph{et al.} 2015 \upcite{Du2015} and the rate of $^7$Be$(e,\nu)^7$Li from the JINA database. The convective overturn-time of the envelope is set to $\sim$1 yr, which is a good approximation in the sense that the mixing between the processed material and the convective envelope is instantaneous. The abundance of the $i$-th nucleus changes in the envelope during the transport is due to the nucleosynthesis and the material replacement, and this corresponds to integrating $\dot{Y}^E_i=(\dot{M}/M_E)(Y^P_i-Y^E_i(0))$\upcite{Nollett2003}, where $M_E$ is the mass of convective envelope. In order to obtain the RGB stellar structure, including in particular the initial abundance composition in the envelope and the distributions of temperature and density in the radiative zone, we calculated the stellar evolution model for $M=1.36\,M_{\odot}$ by the MESA code. The initial abundance of $^7$Li in the sample material at the base of the envelope is $\sim 1.024\times 10^{-11}$ (A(Li)=1.86) obtained from the present stellar evolutionary model calculation, and the $^7$Li abundance may increase to a level of A(Li) exceeding 4 in the processed material when the mass circulation finishes.

During the downward mass circulation, the abundance of $^7$Be increases quickly because the construction reaction $^3$He$(^4$He$,\gamma)^7$Be wins against the destruction reaction $^7$Be$(p,\gamma)^8$B, and the maximum yield of $^7$Be is around the turning point of the mass circulation, where the temperature reaches the highest value $T_p$. In contrast, the production change of $^7$Li behaves rather dramatically  during the downward mass circulation due to the complex competition between the production reaction $^7$Be$(e^-,\nu)^7$Li and the destruction reactions $^7$Li$(p,\gamma)^8$Be and $^7$Li$(p,\alpha)^4$He. The abundance of $^7$Li drops suddenly at about 200 yrs of the processing time as the rates of destruction reactions increase quickly with increasing temperature, and the $^7$Li abundance keeps very low during the downward mass circulation although it increases slightly after about 230 yrs due to the decay of fast growing abundant $^7$Be. In contrast, the abundance of $^7$Li increases sharply during the upward mass circulation due to the fast decreasing of the destruction reaction rates of $^7$Li as the temperature decreases quickly. The processed abundance $Y^P_{Li}$ finally reaches a super-high-level of A(Li)$=4.506$. The abundance of $^7$Li in the envelope contains the contributions of the mass circulation and mass replacement processes, and the total processing time can be estimated as $M_E/\dot{M}$, which is about $2.1\times10^4$ yrs by using the average value of the envelope mass for this RGB star.

\vspace{10pt}
\noindent\textbf{The Projected Rotational Velocity.} Following the assumption of Bruntt \emph{et al.}\upcite{Bruntt2015}, the external broadening of the line profile was assumed to be contributed from the stellar rotation, instrumental broadening, and macro-turbulence. The projected rotational velocity ($v\sin i$) was derived by using five isolated iron lines at 6151\,\AA, 6229\,\AA, 6380\,\AA, 6703\,\AA, and 6810\,\AA. The instrumental broadening was calculated from fitting the emission lines of the arc lamp with a Gaussian profile. The macro-turbulence velocity was estimated using the relation of Hekker \emph{et al.}\upcite{Hekker2007}, which is a function of \Teff\ and \logg. Then we calculated a set of the theoretical spectra broadened with different rotational velocities, and $v\sin i$  was determined by finding the best-fit to the observed iron line profiles.\\


\clearpage
\small
\section*{Supplementary Information}
\setcounter{table}{0}
\setcounter{figure}{0}
\captionsetup[figure]{labelfont={bf},name={Supplementary Figure},labelsep=period}
\captionsetup[table]{labelfont={bf},name={Supplementary Table},labelsep=period}

\begin{table}[!h]
\begin{center}
\caption{Other information of TYC\,$429$-$2097$-$1$}\label{s_tab1}
\begin{tabular}{rcl}
\hline\hline\noalign{\smallskip}
 Property & & Value\\
\noalign{\smallskip}
\hline\noalign{\smallskip}
 Position (J2000)          & R.A.           & $17:53:46.07$   \\
                           & DEC.           & $06:42:41.20$   \\
                           & $l$            & $32.4111$       \\
                           & $b$            & $15.8647$       \\
 \noalign{\smallskip}
 \hline
 \noalign{\smallskip}
 $V_{\rm mag}$             & (mag)          & $11.27$ (UCAC4) \\
 Gaia parallax $\pi$       & (milli-arcsec) & $0.73\pm0.24$   \\
 $A_{\rm V}$               & (mag)          & $0.47$          \\
 $BC$\                     & (mag)          & $-0.385$        \\
 $M_{\rm bol}$             & (mag)          & $-0.27$         \\
 \noalign{\smallskip}
 \hline
 \noalign{\smallskip}
 $[$C/Fe$]$                &                & $-0.02\pm0.10$  \\
 $[$N/Fe$]$                &                & $ 0.45\pm0.10$  \\
 $[$Mg/Fe$]$               &                & $ 0.23\pm0.05$  \\
 $[$Si/Fe$]$               &                & $ 0.19\pm0.01$  \\
 $[$Ca/Fe$]$               &                & $ 0.16\pm0.06$  \\
 \noalign{\smallskip} \hline
 \end{tabular}
 \end{center}
 \end{table}

\begin{table}[!h]
\begin{center}
\caption{The variation of Li abundance caused by the uncertainties of stellar parameters }\label{s_tab2}
\begin{tabular}{cccc}
\hline\hline\noalign{\smallskip}
 Wavelength & $\Delta$\Teff & $\Delta \log g$ & $\Delta$[Fe/H] \\
\noalign{\smallskip}
            & $80$\,K       & $0.10$\,dex   & $0.06$\,dex    \\
\noalign{\smallskip}
\hline\noalign{\smallskip}
 $6103.6$\,\AA  & $0.09$ & $0.02$ & $0.07$ \\
 $6707.8$\,\AA  & $0.09$ & $0.03$ & $0.06$ \\
 $8126.3$\,\AA  & $0.08$ & $0.01$ & $0.07$ \\
 \noalign{\smallskip} \hline
 \end{tabular}
 \end{center}
 \end{table}

\begin{figure}[!h]
\setlength{\abovecaptionskip}{-15pt}
\setlength{\belowcaptionskip}{-15pt}
\includegraphics[angle=0, width=\hsize]{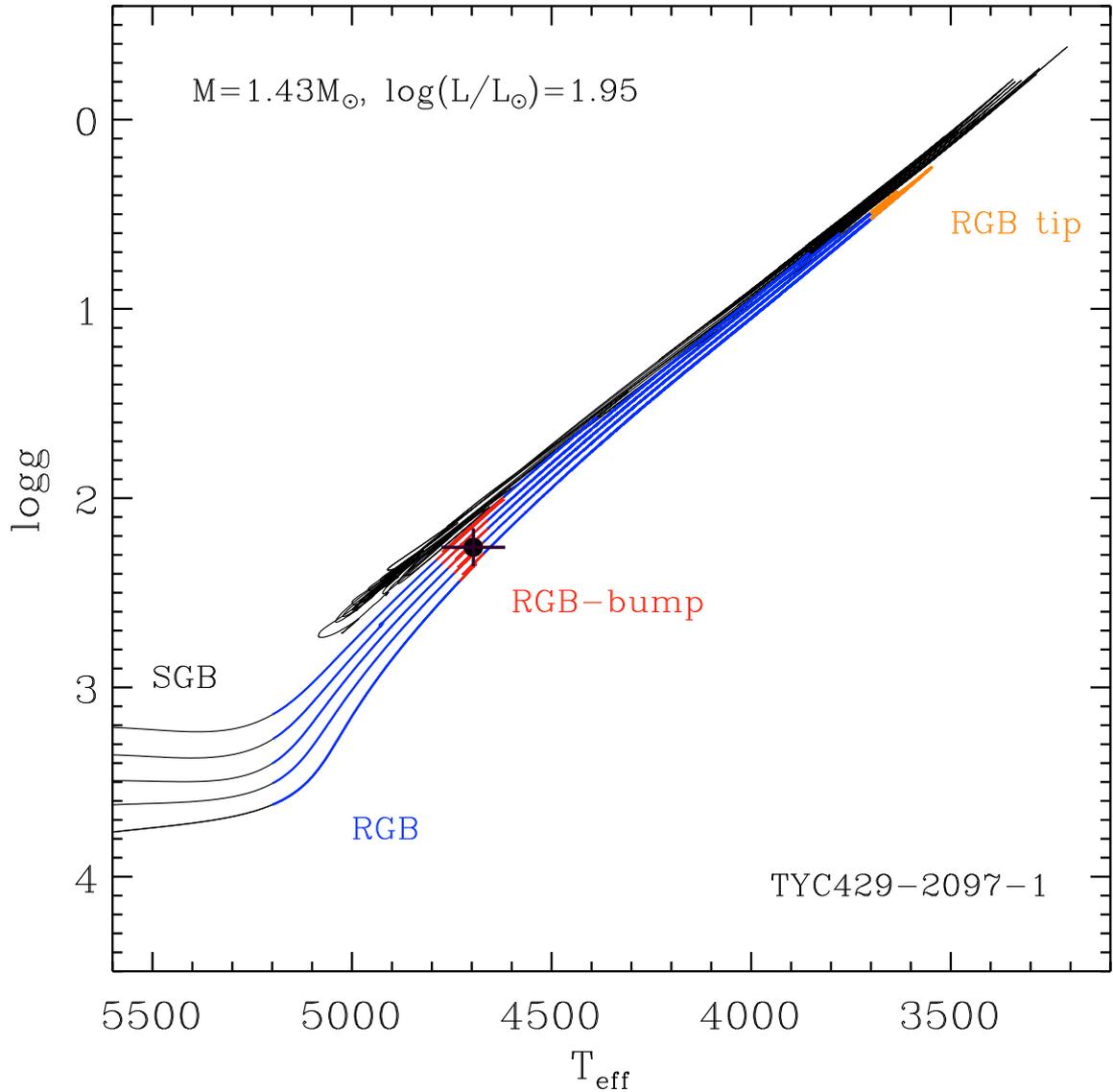}\\
\caption{The corresponding evolutionary tracks computed from the MESA code for TYC\,$429$-$2097$-$1$. The mass range of the tracks is from $1.0\,M_{\odot}$ (bottom) to $1.8\,M_{\odot}$ (top), with an interval of $0.2\,M_{\odot}$. The evolutionary stages for each track are indicated with different colors, namely: `black' $-$ SGB, `dark blue' $-$ RGB, `red' $-$ RGB-bump and `orange' $-$ RGB tip. The position of TYC\,$429$-$2097$-$1$ is indicated with a black dot. The error bars indicate the uncertainties in \Teff\ and $\log g$, which are adopted to be $\pm 80$\,K and $0.10$\,dex from our stellar parameters determination method.}\label{s_fig1}
\end{figure}

\begin{figure}[!h]
\setlength{\abovecaptionskip}{-10pt}
\setlength{\belowcaptionskip}{-10pt}
\includegraphics[angle=0,width=\hsize]{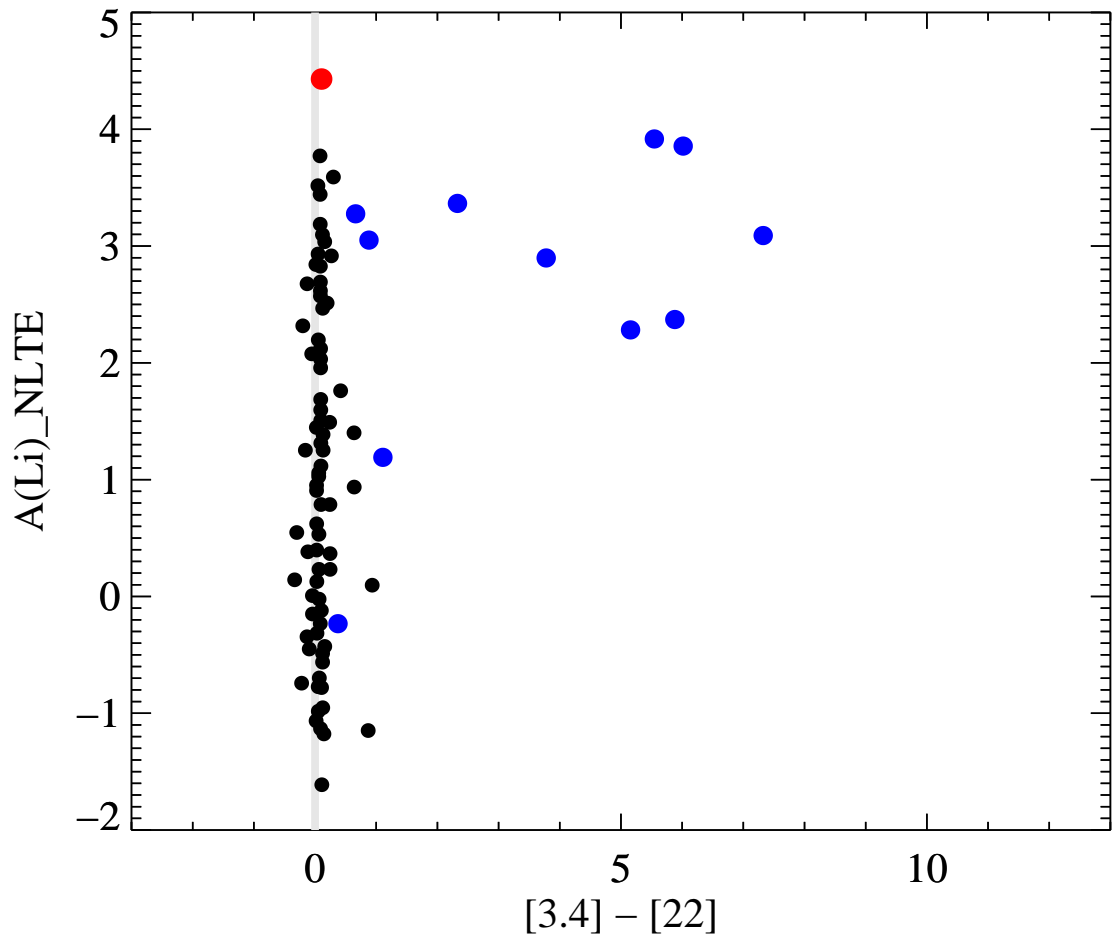}\\
\caption{The IR-excess diagram for TYC\,$429$-$2097$-$1$. TYC\,$429$-$2097$-$1$ is is indicated with a red dot, while the other stars are from Rebull {\emph et al.}\upcite{Rebull2015} 2015. The horizontal line at A(Li) = $1.5$ shows the adopted division between Li-rich and normal stars, and the vertical line at $[3.4]-[22]=0$ indicates the photospheric locus. The blue dots denote the objects with strong IR-excess, while the black dots represent the stars with no significant IR-excess.}\label{s_fig2}
\end{figure}

\begin{figure}[!h]
\setlength{\abovecaptionskip}{-10pt}
\setlength{\belowcaptionskip}{-10pt}
\includegraphics[angle=0,width=\hsize]{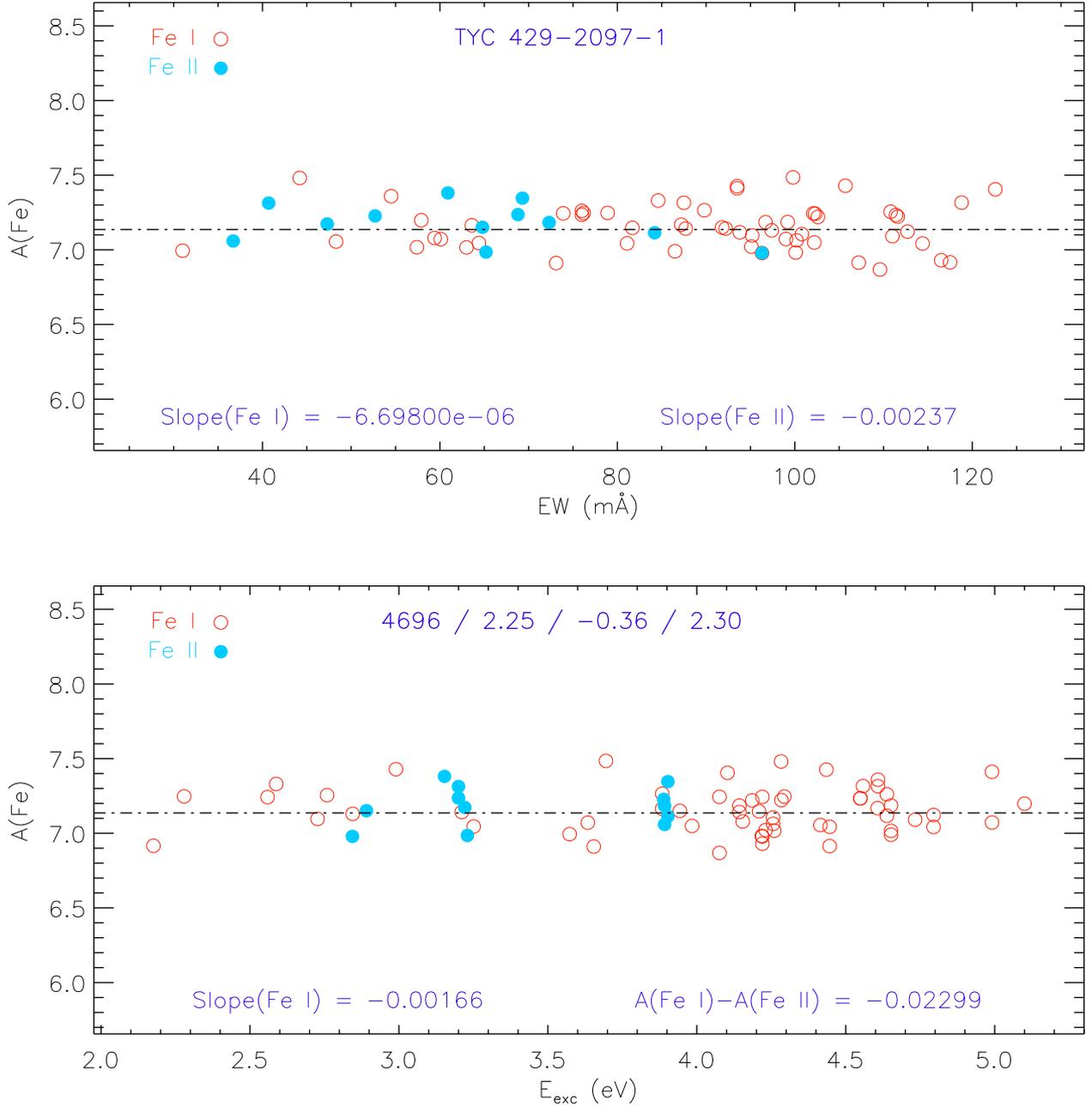}\\
\caption{The determination of the stellar parameters. The figure shows the absolute NLTE abundance from lines of \ion{Fe}{1} (red open circle) and \ion{Fe}{2} (blue dots) in TYC\,$429$-$2097$-$1$ as functions of their EWs (top panel) and $E_{\rm exc}$ (bottom panel). The slopes are indicated in the corresponding panel. The mean A(Fe) averaged from \ion{Fe}{1} and \ion{Fe}{2} lines is shown with a dash-dotted line in both panels. The final stellar parameters are denoted in the seqence of \Teff\,(K), $\log g$, [Fe/H], and $\xi_t$ (km\,s$^{-1}$) in the bottom panel.}\label{s_fig3}
\end{figure}

\end{document}